\documentclass[onecolumn]{article}
\usepackage{epsfig}
\usepackage{multicol}
\usepackage{array}
\usepackage{amsfonts,amsmath,amssymb}

\input{epsf.sty}
\begin{document}
\begin{center}
{\Large{\bf Radiative transfer in scattering stochastic atmospheres}}\\
\bigskip
{N. A. Silant'ev\thanks{E-mail: nsilant@bk.ru}\,\,, G. A. Alekseeva \,,\, V. V. Novikov}
\medskip\\
\end{center}
\begin{center}
 {Central Astronomical Observatory at Pulkovo of Russian Academy of Sciences,\\ 196140,
Saint-Petersburg, Pulkovskoe shosse 65, Russia\\}
\medskip
\end{center}

\bigskip
\begin{center}
{received..... 2017, \qquad accepted....}
\end{center}

\begin{abstract}
Many stars, active galactic nuclei, accretion discs etc. are affected by  the stochastic variations of temperature, turbulent gas motions, magnetic fields, number densities of atoms and dust grains. These stochastic variations  influence on the extinction factors, Doppler  widths of lines and so on. The presence of many reasons for fluctuations gives rise  to Gaussian distribution of fluctuations. The usual models  leave out of account the fluctuations. In many cases the consideration of fluctuations  improves the coincidence of theoretical values with the observed data. The objective of this paper is the investigation of the influence of the number density fluctuations on  the form  of radiative transfer equations. We consider non-magnetized atmosphere in continuum.  

\end{abstract}

$^1$ E-mail: nsilant@bk.ru

\section{Introduction}
In this paper we consider the derivation of radiative transfer equations in scattering stochastic atmospheres. The ensemble of fluctuations of the extinction factor $\alpha=N\sigma_t$ and the scattering coefficient  $\alpha_s=N\sigma_s$  is assumed to be of Gaussian type.
Here $N$ is the number density of scattering particles; $\sigma_t=\sigma_s +\sigma_a$, where $\sigma_s$ and $\sigma_a$ are the scattering cross-section and the absorption cross-section, respectively.

Thus, we consider the  stochastic atmosphere, where  the fluctuating values $N'$  are number density fluctuations  of  the scattering particles. The reasons of fluctuations are the  compressible turbulence or the temperature fluctuations, which also give rise to fluctuations of the number density.  Both reasons really hold in star's atmospheres and  accretion discs around Active Galactic Nuclei (AGN). For example, in our Sun the temperature fluctuations $\Delta T/T\sim 0.02-0.03$ are valid (see Stix 1991). Recall, that for adiabatic  processes  the relation $N_1/N_2=(T_1/T_2)^{\gamma-1}$ holds. The value $\gamma =5/3$ corresponds to monoatomic gas and for diatomic gas $\gamma=7/2$ .  This relation means that the temperature fluctuations give rise to the fluctuations of the number densities: $\Delta N_1/N_1+\Delta N_2/N_2\simeq (\gamma-1)(\Delta T_1 /T_1+ \Delta T_2/T_2)$. Here we assumed that the temperature fluctuations are small $\Delta T/T\ll 1$.    No doubt, the temperature fluctuations  hold in different types of stars.  It is apparently that the most large fluctuations $N'$ deal with the shock waves turbulence. 

 It is known that Gaussian ensemble  corresponds to  many chaotic reasons to arise the fluctuations. The observed Stokes parameters  are the temporal and space averaged values. The ensemble averaging we denote by $\langle \rangle$. The averaged value $\langle I\rangle $ is denoted by $I_0$. The fluctuations are denoted by primes $I'$ ($\langle I' \rangle=0)$.

 We use the standard formulas for the averaging of stochastic values (see van Kampen 1981, Gardiner 1985). For convenience of readers we at first present them.

 For Gaussian ensemble of realizations the average of the odd number of fluctuating quantities is equal to zero and the average of the even numbers is equal to the sum  of all possible two-point correlation functions (correlators) (see van Kampen 1981, Gardiner 1985). Thus, for example:
\begin{equation}
\langle x_1 x_2 x_3 x_4 \rangle=\langle x_1x_2\rangle \langle x_3x_4\rangle +\langle x_1x_3\rangle \langle x_2x_4\rangle +
\langle x_1x_4\rangle \langle x_2x_3\rangle \to \langle x^4\rangle=3\langle x^2\rangle ^2,
\label{1}
\end{equation}
\noindent where $x_i$ are fluctuating values with zero mean value $\langle x_i\rangle=0$.  In right side of Eq.(1) are 3 terms. The average of the value $x_1x_2x_3x_4x_5x_6$ can be written as:
\[
\langle x_1x_2x_3x_4 x_5 x_6 \rangle=\langle x_1x_2\rangle \langle x_3x_4 x_5 x_6\rangle +
\langle x_1x_3 \rangle \langle x_2 x_4 x_5 x_6 \rangle +
\]
\begin{equation}
\langle x_1x_4 \rangle \langle x_2x_3 x_5 x_6 \rangle 
+ \langle x_1x_5\rangle \langle x_2x_3 x_4 x_6 \rangle +\langle x_1x_6\rangle \langle x_2x_3 x_4 x_5 \rangle \to 5\langle x^2\rangle\langle x^4\rangle= 5\cdot 3\langle x^2\rangle^3.
\label{2}
\end{equation}
\noindent Eq.(2) , according to Eq.(1), consists of $5\times 3=15$ terms.  The arrow shows the transition to the case when all $x_i$ are equal to $x$. For this case we have the recurrence formulas:
\begin{equation}
\langle x^{2k}\rangle =(2k-1)\langle x^2\rangle \langle x^{(2k-2)}\rangle= (2k-1)(2k-3)(2k-5)....1\langle x^2\rangle^k\equiv (2k-1)!!\langle x^2\rangle^k,
\label{3}
\end{equation}

\begin{equation}
\langle ax^{2k+1}\rangle =(2k+1)\langle ax\rangle \langle x^{2k}\rangle =\langle ax\rangle (2k+1)!!\langle x^2\rangle^k.
\label{4}
\end{equation}
\noindent Here $a$ is  Gaussian fluctuation with $\langle a\rangle=0$.
\noindent Recall, that standard factorial $n!=n(n-1(n-2)...1$, i.e. $n!=n(n-1)!$. The factorial consisting of the odd numbers is denoted as $(2n-1)(2n-3)(2n-5)...1\equiv (2n-1)!!$. This value obeys the recurrence formula  $(2n-1)!!=(2n-1)(2n-3)!!$. It is easy prove the
formula:$ (2n)!=2^n n!(2n-1)!!.$

The mean values of $\exp{(\pm\tau')}$ and $a'\exp{(-\tau')}$ can be readily obtain, using the average  procedure,  given in Eqs.(1) - (4), for the standard  presentation  of the exponential function $\exp{(x)}$:
\begin{equation}
\exp{(\pm x)}=1+\sum_{k=1}^{\infty}\frac{x^{2k}}{(2k)!} \pm \sum_{k=0}^{\infty}\frac{x^{2k+1}}{(2k+1)!}.
\label{5}
\end{equation}
\noindent Using the recurrence formulas (3) and (4), we obtain the following expressions:
\begin{equation}
\langle \exp{(\pm\tau')}\rangle =\exp{\left(\frac{\langle \tau'^2 \rangle }{2}\right)},
\label{6}
\end{equation}
\begin{equation}
\langle a' \exp{(-\tau')}\rangle =-\langle a'\tau' \rangle \exp{\left(\frac{\langle \tau'^2 \rangle }{2}\right)}.
\label{7}
\end{equation}
\noindent Here $\tau'$ - values are fluctuations of an optical depth.

  Formulas (6) and (7) were given in Silant'ev (2005) without derivation. In this paper the propagation of polarized radiation in  magnetized electron atmosphere is considered, i.e. the free electrons  was taken as the only scattering particles.

 Below we consider the radiative transfer equation in scattering stochastic atmospheres both consisting of one type scattering particles and  two types ones. We consider also the matrix transfer equation for Stokes parameters  $I$ and $Q$ in the atmosphere, consisting of free electrons and small  anisotropic particles (molecules or dust grains).  We do not take into account the possible magnetic field in an atmosphere.   Note that the new extinction factor $\alpha_{eff}<\alpha$ arises in all atmospheres.
In two-component atmosphere, the radiative transfer equation takes the new form for integral terms.

Firstly we consider the transfer equation for the mean intensity $\langle I({\bf n},r)\rangle\equiv I_0({\bf n},r)$ in  stochastic atmosphere, consisting of one type particles.  Note that we use the notations  $\langle \alpha\rangle \equiv \alpha_0 $ and the averaged optical depth $\langle\tau\rangle \equiv \tau_0$.

\section{Stochastic atmosphere with one type of scattering particles}

 In stochastic atmosphere, the extinction factor has a fluctuating component: $\alpha=\langle\alpha\rangle+\alpha'\equiv \alpha_0(r)+\alpha'(r)$, $\langle \alpha'( r)\rangle=0$. The change of radiation intensity along the path $r$ is determined by the  standard differential  equation:
\begin{equation}
dI({\bf n},r)=-[\alpha_0(r)+\alpha'(r)]I({\bf n},r)dr\equiv-[N_0(r)+N'(r)]\sigma_t I({\bf n},r)dr.
\label{8}
\end{equation}
\noindent Here $\alpha(r)=N(r)\sigma_t$, where $N(r)=N_0(r) +N'(r)$ is the number density of scattering particles. The value $\sigma_t=\sigma_s+\sigma_a$ is the total
cross-section, which is the sum of scattering cross-section $\sigma_s $ and  absorption cross-section $\sigma_a$.   The  differentiation  shows that the solution of Eq.(8) is:
\begin{equation}
I({\bf n},r)=I({\bf n},0)\exp{\left[-\int_0^rdr'(\alpha_0(r')+\alpha'(r'))\right]}\equiv
 I({\bf n},0)\exp{[-(\tau_0(r)+\tau'(r))]},
\label{9}
\end{equation}
\noindent where $\tau_0(r)$ and $\tau'(r)$ are:
\begin{equation}
\tau_0(r)=\int^r_0dr'\,\alpha_0(r')=\sigma_t\int^r_0dr'\,N_0(r'),
\label{10}
\end{equation}
\begin{equation}
\tau'(r)=\int^r_0dr'\,\alpha'(r')=\sigma_t\int^r_0dr'\,N'(r').
\label{11}
\end{equation}
\noindent The averaging of Eq.(9)  with allowance  for Gaussian distribution  of fluctuations probability   (see Eq.(6)) gives rise to:
\begin{equation}
\langle I({\bf n},r)\rangle\equiv I_0({\bf n},r)=I({\bf n},0)\exp{\left[-\left(\tau_0(r)-\frac{1}{2}\langle\tau'^2(r)\rangle\right)\right]}
\equiv I({\bf n},0)\exp{[-\tau_{eff}(r)]},
\label{12}
\end{equation}
\begin {equation}
\tau_{eff}(r)=\tau_0(r)-\frac{1}{2}\langle\tau'^2(r)\rangle.
\label{13}
\end{equation}
\noindent Eq.(12) shows that the averaged intensity $I_0({\bf n},r) $ in stochastic medium decreases
with the distance weaker than intensity with accounting for the mean absorption factor $\alpha_0(r)$.

 Let us demonstrate this by the averaging of two realizations:

\[
I_1({\bf n},r)=I({\bf n},0)\exp{(-\tau_0(r)+\tau'(r))},
\]
\begin{equation}
I_2({\bf n},r)=I({\bf n},0)\exp{(-\tau_0(r)-\tau'(r))}.
\label{14}
\end {equation}
\noindent The mean value of these two realizations is:
\begin{equation}
I_0({\bf n},r)=\frac{I_1({\bf n},r) +I_2({\bf n},r)}{2}=
I({\bf n},0)\exp{(-\tau_0(r))}\cosh{(\tau'(r))}.
\label{15}
\end{equation}
\noindent We see that the mean value is  larger than $ I({\bf n},0)\exp{(-\tau_0(r))}$ ($\cosh{(x)}\ge1$)  . This is the statistical effect.

Using Eq.(7), one can derive  the following expression:
\begin{equation}
\langle \alpha(r) I({\bf n},r)\rangle=\alpha_0(r) I_0({\bf n},r) +\langle\alpha'(r) I'({\bf n},r)\rangle=
[\alpha_0(r)-\langle\alpha'(r)\tau'(r)\rangle] I_0({\bf n},r)\equiv\alpha_{eff}(r)I_0({\bf n},r),
\label{16}
\end{equation}
\noindent where
\begin{equation}
\alpha_{eff}(r)=\alpha_0(r)-\langle\alpha'(r)\tau'(r)\rangle.
\label{17}
\end{equation}

The standard radiative transfer equation for $ I({\bf n},r)$  has the form (see Chandrasekhar 1960):

\begin{equation}
\frac{d I({\bf n},r)}{dr}=({\bf n}\nabla)I({\bf n},r)=-\alpha(r) I({\bf n},r) +
\frac{\alpha^{(s)}(r)}{4\pi}\int d{\bf n'}\,p({\bf n\cdot n'}) I({\bf n'},r) +S({\bf n},r).
\label{18}
\end{equation}

 Here $p({\bf n\cdot n'})$ is the phase function, $ S({\bf n},r)$ is the source function.
Note that the extinction factor $\alpha(r)=\alpha^{(s)}(r)+\alpha^{(a)}(r)$ describes the extinction  due to scattering and pure absorption, respectively. Recall, that  $\alpha^{(s)}(r) =N(r)\sigma_s$ and $\alpha^{(a)}(r)=N(r)\sigma_a$. Here $\sigma_s$  and $\sigma_a$ are the cross-sections of scattering and absorption, respectively. The total cross-section $\sigma_t=\sigma_s+\sigma_a$. The degree of absorption $\varepsilon=\sigma_a/\sigma_t$. The extinction factor $\alpha(r)=N(r)\sigma_t$.   The value $N(r)$ and, consequently, $\alpha(r)$ has the fluctuating component. It is clear that  $\alpha^{(s)}(r)\equiv(1-\varepsilon)\alpha(r)$.

The phase function  $p({\bf n\cdot n'})$  depends on the scalar product of vectors ${\bf n'}$ and ${\bf n}$ - directions of incident radiation and the observed direction. Integral of  $p({\bf n\cdot n'})/4\pi$ over  ${\bf n}$ or ${\bf n'}$ is equal to unity. In conservative atmosphere ($\varepsilon=0$) isotropic phase function is  $p({\bf n\cdot n'})=1$ and for Rayleigh scattering  $p({\bf n\cdot n'})=3/4(1+{\bf n\cdot n'})$.

 It is easy prove, that in conservative atmosphere ($\alpha_{a}=0,  \varepsilon=0$) and the source $S({\bf n},r)=0$  the law of conservation of radiation flux exists:

\begin{equation}
\frac{{d\bf F}(r)}{dr}=0,  \,\,\, {\bf F}(r)=\int _{(4\pi)}d{\bf n}\,{\bf n}I({\bf n},r).
\label{19}
\end{equation}
\noindent  The derivation of the conservation law  uses the property:
\begin{equation}
\frac{1}{4\pi}\int _{(4\pi)}d{\bf n}\,p({\bf n\cdot n'})=1.
\label{20}
\end{equation}
\noindent  The conservation law follows from physical consideration.  Eq.(19) confirms that the radiative transfer equation is true.

The averaging of Eq.(18), with the taking into account the relation (16), gives rise to the radiative transfer equation for the mean intensity $\langle I({\bf n},r)\rangle \equiv  I_0({\bf n},r)$:
\begin{equation}
\frac{d I_0({\bf n},r)}{dr}=({\bf n}\nabla)I_0({\bf n},r)=-\alpha_{eff}(r)  I_0({\bf n},r)+
\frac{ \alpha^{(s)}_{eff}(r) }{4\pi}\,\int d{\bf n'}\,p({\bf n\cdot n'}) I_0({\bf n'},r)
  +\langle S({\bf n},r)\rangle,
\label{21}
\end{equation}
\noindent where
\begin{equation}
 \alpha^{(s)}_{eff}=\alpha^{(s)}_0(r)- \langle \alpha'^{(s)}(r)\tau'(r)\rangle\equiv(1-\varepsilon)\alpha_{eff}(r).
\label{22}
\end{equation}
\noindent  

 Thus, using the supposition that the stochastic atmosphere physically presents  the Gaussian ensemble of realizations, we obtained the
closed radiative transfer  equation for mean intensity $ I_0({\bf n},r)$. The main role in this presentation plays  Eq.(16). 

 The relation (16) takes place
for radiation going through the stochastic Gaussian type  atmosphere. Clearly, in the integral term of Eq.(18) the intensity also arises after some transportation through the stochastic medium. Note that  the conservation law (Eq.(19)) holds also for  Eq.(21)  (recall, that in conservative atmosphere $\varepsilon=0$ and $\alpha_{eff}=\alpha^{(s)}_{eff}$).

 The transfer equation in stochastic atmosphere, where $d\tau=\alpha_{eff}ds$, takes the standard form:
\begin{equation}
\frac{d I_0({\bf n},\tau)}{d\tau}=({\bf n}\nabla_{\tau})I_0({\bf n},\tau)=- I_0({\bf n},\tau) +
\frac{1-\varepsilon}{4\pi}\int d{\bf n'}\,p({\bf n\cdot n'}) I_0({\bf n'},\tau) +S_0({\bf n},\tau).
\label{23}
\end{equation}
\noindent It is interesting to note that the form of transfer equation for the mean intensity  is similar to standard radiative transfer equation with $d\tau=\alpha_0 dr$.

  Because of $\alpha_{eff}<\alpha_0$ , the geometrically similar layers have different optical depths - the stochastic layer has effectively smaller (more transparent) depth than non-stochastic one. Eq.(21) uses the  supposition that the ensemble of fluctuations is Gaussian.

 Now we consider the averaged quantities, assuming
that two-point correlations like $\langle \tau'^2\rangle$ etc., depend on $|r'-r''|=R$, i.e. we assume the  model of homogeneous and isotropic turbulence. Let us consider the case  $\langle \tau'^2\rangle$ in more detail:
\[
\langle \tau'^2\rangle=\int_0^rdr'\int_0^rdr''\langle \alpha'(r')\alpha'(r'')\rangle\equiv
\]
\begin{equation}
\langle \alpha'^2\rangle \int_0^rdr'\int_0^rdr'' A_{\alpha}(R/R_1)\equiv
2\langle \alpha'^2\rangle\int_0^r\,dR(r-R)A_{\alpha}(R/R_1).
\label{24}
\end{equation}
\noindent Here we introduced  two-point correlation function  $A_{\alpha}(R/R_1)$:
\begin{equation}
\langle \alpha'(r')\alpha'(r'')\rangle=\langle \alpha'^2\rangle A_{\alpha}(R/R_1)=\langle N'^2\rangle \sigma_t^2 A_{\alpha}(R/R_1).
\label{25}
\end{equation}
\noindent The  function $A_{\alpha}(R/R_1)$ is equal to unity at $R=0$. The parameter $R_1$ is the length of correlation, i.e. for $R>R_1$ correlation between $\alpha'(r')$ and $\alpha'(r'')$ tends to zero. 
The value $\langle \alpha'^2\rangle $ gives the mean value of $\langle \alpha'(r')\alpha(r'')\rangle$ for $r'=r''$.

Frequently one considers the homogeneous medium with $\alpha_0(r)=\alpha_0$. Below we  restrict ourselves by this case. In this case Eq.(24) can be written in the form:
\begin{equation}
\langle \tau'^2\rangle=2\frac{\langle \alpha'^2\rangle}{\alpha^2_0}\tau_0\tau_1 f_{\alpha}(r/R_1)\equiv2\frac{\langle N'^2\rangle}{N_0^2}\tau_0\tau_1f_{\alpha}(r/R_1).
\label{26}
\end{equation}
\noindent Here $\tau_0=\alpha_0\,r$ and $\tau_1=\alpha_0\,R_1$ are the total mean optical depths of the distance $ r$  and the correlation length $R_1$, respectively.  The value $f_{\alpha}(r/R_1)$ denotes the integral:
\begin{equation}
f_{\alpha}(r/R_1)=\int_0^{r/R_1}dx \left(1-x\frac{R_1}{r}\right)A_{\alpha}(x).
\label{27}
\end{equation}

The characteristic length $r_0$ corresponds to  free path.
If the length $r_0>>R_1$ , then the value $f_{\alpha}$ tends to the simple expression:
\begin{equation}
f_{\alpha}(r/R_1)\to\int_0^{\infty}dx\,A_{\alpha}(x)\equiv f_{\alpha}.
\label{28}
\end{equation}
\noindent Frequently one assumes $A_{\alpha}(x)=\exp{(-x)}$. In this case $ f_{\alpha}=1$. For other peak-like forms of correlation function $A_{\alpha}(x)$ we have  $ f_{\alpha}\simeq 1$. Thus, for estimate of $\langle \tau'^2\rangle $ one can use $ f_{\alpha}\simeq 1$. Using Eq.(26),  we obtain for $\tau_{eff}$ the value:

\begin{equation}
\tau_{eff}(r)=\tau_0+\frac{1}{2}\langle \tau'^2\rangle=\tau_0\left(1-\frac{\langle \alpha'^2\rangle}{\alpha^2_0}\tau_1 f_{\alpha}\right).
\label{29}
\end{equation}

For correlation function $\langle \alpha'(r)\tau'(r)\rangle$   we obtain the formula:

\begin{equation}
\langle \alpha'(r)\tau'(r)\rangle =\frac{\langle \alpha'^2\rangle\tau_1}{\alpha_0}\int_0^{r/R1}dxA_{\alpha}(x)\to
\frac{\langle \alpha'^2\rangle\tau_1}{\alpha_0}\int_0^{\infty}dxA_{\alpha}(x)\simeq \frac{\langle \alpha'^2\rangle\tau_1}{\alpha_0}f_{\alpha}.
\label{30}
\end{equation}
\noindent 
\noindent As a result, Eq.(17) for $\alpha_{eff}(r)$ takes the form:
\begin{equation}
  \alpha_{eff}(r)=[\alpha_0(r)-\langle \alpha'(r)\tau'(r)\rangle ] \to
 \alpha_0\left(1-\frac{\langle \alpha'^2\rangle}{\alpha^2_0}\tau_1 f_{\alpha}\right)\equiv \alpha_0\left(1-\frac{\langle N'^2\rangle}{N_0^2}\tau_1f_{\alpha}\right).
\label{31}
\end{equation}

 We consider the case when the correlation optical  length  $\tau_1=\alpha_0 R_1<<1$. The level of fluctuations $\langle \alpha'^2\rangle/\alpha^2_0$, in principle, may be arbitrary with the only restriction $\alpha_{eff}>0$. It appears  most large
fluctuations $N'$ hold in a chaotic shock waves turbulence.  Recall (see Eq.(22)), that $\alpha^{(s)}_{eff}=(1-\varepsilon)\alpha_{eff}$.

 It is known that emerging radiation  mainly goes from optically thickness $\sim 1$. In stochastic atmosphere, where  $\alpha_{eff}<\alpha_0$ , the emerging radiation goes from deeper layers, than in  non-stochastic one. Usually in deep layers the temperature is larger. Thus, the stochastic atmosphere demonstrates more larger temperature than non-stochastic one.

Note that radiation scattering on various types of particles (say, with $p_1({\bf nn'})$  and  $p_2({\bf nn'})$ ) in stochastic atmosphere
changes the form of transfer equation more significant. This will be clear in the next section, where we consider the scattering on two types of particles. 

\section{Stochastic atmosphere with two type of scattering particles}

In previous section we considered the stochastic atmosphere with the identical scattering particles.  Now we consider more complex situation when the stochastic atmosphere consists of two types of scattering particles. As an example of such atmosphere can be considered   the gas dusty accretion disc  and the torus around the AGNs nuclei.

Let these components are characterized by the number densities $N_1(r)=N_1^{(0)}+N'_1$ and $N_2(r)=N_2^{(0)}+N'_2$. The corresponding cross-sections are $\sigma_1^{(t)}=\sigma_1^{(s)}+\sigma_1^{(a)}$ and $\sigma_2^{(t)}=\sigma_2^{(s)}+\sigma_2^{(a)}$. The standard radiative transfer equation in this case has the form:
\[
\frac{d I({\bf n},r)}{dr}=({\bf n}\nabla)I({\bf n},r)=-\alpha I({\bf n},r) +
\frac{\alpha^{(s)}_1}{4\pi}\int d{\bf n'}\,p_1({\bf n\cdot n'}) I({\bf n'},r)+
\]
\begin{equation}
\frac{\alpha^{(s)}_2}{4\pi}\int d{\bf n'}\,p_2({\bf n\cdot n'}) I({\bf n'},r)  +S({\bf n},r),
\label{32}
\end{equation}
\noindent where
\[
 \alpha\equiv\alpha_0= N_1^{(0)}\sigma^{(t)}_1+N_2^{(0)}\sigma^{(t)}_2\equiv \alpha^{(0)}_1+\alpha^{(0)}_2,
\]
\[
 \alpha^{(s)}_1=N_1^{(0)}\sigma^{(s)}_1=(1-\varepsilon_1)\alpha^{(0)}_1, 
\]
\begin{equation}
  \alpha^{(s)}_2=N_2^{(0)}\sigma^{(s)}_2=(1-\varepsilon_2)\alpha^{(0)}_2. 
\label{33}
\end{equation}

  It is easy prove that the conservation law of the radiative flux (see Eq.19) holds in this transfer equation. This law corresponds to conservative  atmosphere without the source of radiation, i.e. $\varepsilon_1=0, \varepsilon_2=0, S({\bf n}, r)=0.$ Recall, that the phase functions $p_1({\bf n}\cdot {\bf n'})$ and $p_2({\bf n}\cdot {\bf n'})$ obey the condition (20).

If we assume $d\tau=\alpha_0 dr =(N_1^{(0)}\sigma^{(t)}_1+N^{(0)}_2\sigma^{(t)}_2)dr\equiv N^{(0)}_2\sigma^{(t)}_2(1+\eta)dr$, then Eq.(32) takes the form:

\[
\frac{d I({\bf n},\tau)}{d\tau}=({\bf n}\nabla_{\tau})I({\bf n},\tau)=- I({\bf n},\tau) +
\frac{(1-\varepsilon_1)\eta}{4\pi (1+\eta)}\int d{\bf n'}\,p_1({\bf n\cdot n'}) I({\bf n'},\tau)+
\]
\begin{equation}
\frac{1-\varepsilon_2}{4\pi(1+\eta)}\int d{\bf n'}\,p_2({\bf n\cdot n'}) I({\bf n'},\tau)  +S({\bf n},\tau),
\label{34}
\end{equation}
\noindent  where the parameter $\eta$ is equal to:
\begin{equation}
\eta=\frac{N^{(0)}_1\sigma^{(t)}_1}{N^{(0)}_2\sigma^{(t)}_2}\equiv \frac{\alpha^{(0)}_1}{\alpha^{(0)}_2}.
\label{35}
\end{equation}
\noindent  Thus, Eq.(34) depends on one parameter $\eta$. 

Assuming that all factors in Eq.(32)  are stochastic values, we obtain the following transfer equation for the averaged  intensity $I_0({\bf n},r)$:
\[
\frac{d I_0({\bf n},r)}{dr}=({\bf n}\nabla)I_0({\bf n},r)=-\alpha_{eff} I({\bf n},r) +
\frac{\alpha^{(s)}_{1eff}}{4\pi}\int d{\bf n'}\,p_1({\bf n\cdot n'}) I_0({\bf n'},r)+
\]
\begin{equation}
\frac{\alpha^{(s)}_{2eff}}{4\pi}\int d{\bf n'}\,p_2({\bf n\cdot n'}) I_0({\bf n'},r)  +S_0({\bf n},r),
\label{36}
\end{equation}
\noindent where
\[
\alpha_{eff}\equiv \alpha_{1eff}+\alpha_{2eff }=[N^{(0)}_1\sigma^{(t)}_1 -(\gamma_1+\gamma_3)]+[N^{(0)}_2\sigma^{(t)}_2 -(\gamma_2+\gamma_3)],
\]

\begin{equation}
 \gamma_1=\langle N'^2_1\rangle ( \sigma^{(t)}_1)^2 R_1f_{1\alpha},\quad \gamma_2=\langle N'^2_2\rangle ( \sigma^{(t)}_2)^2 R_2f_{2\alpha},\quad \gamma_3=\langle N'_1N'_2\rangle \sigma^{(t)}_1\sigma^{(t)}_2 R_3f_{3\alpha}.
\label{37}
\end{equation}
\begin{equation}
\alpha_0=N^{(0)}_1\sigma^{(t)}_1 +N^{(0)}_2\sigma^{(t)}_2,\quad \alpha'=N'_1\sigma^{(t)}_1+N'_2\sigma^{(t)}_2,
\label{38}
\end{equation}

\noindent Here we introduced  the different lengths of correlation and correlation functions for the values $\langle N'_1 N'_1\rangle$, $\langle N'_2 N'_2\rangle$ and $\langle N'_1 N'_2\rangle$. In general case they are  different one from another.

\begin{equation}
\alpha^{(s)}_{1eff}=(1-\varepsilon_1)[N_1^{(0)}\sigma^{(t)}_1- (\gamma_1+\gamma_3)]\equiv(1-\varepsilon_1)\alpha_{1eff},
\label{39}
\end{equation}

\begin{equation}
\alpha^{(s)}_{2eff}=(1-\varepsilon_2)[N_2^{(0)}\sigma^{(t)}_2- (\gamma_2+\gamma_3)]\equiv(1-\varepsilon_2)\alpha_{2eff}.
\label{40}
\end{equation}
\noindent Recall, that $\varepsilon_1=\sigma^{(a)}_1/\sigma^{(t)}_1$ and  $\varepsilon_2=\sigma^{(a)}_2/\sigma^{(t)}_2$ are degrees of absorption of the first and second types of scattering particles, respectively. The values $R_i$ 
 and the values $f_{i\alpha}$  correspond to different correlation functions  (see Eq.(25)).

If we assume the optical depth $d\tau=\alpha_{eff}dr $, then Eq.(36) can be presented in the form of Eq.(34)
 with $\eta \to \eta_{eff}$:
\[
\frac{d I({\bf n},\tau)}{d\tau}=({\bf n}\nabla_{\tau})I({\bf n},\tau)=- I({\bf n},\tau) +
\frac{(1-\varepsilon_1)\eta_{eff}}{4\pi (1+\eta_{eff})}\int d{\bf n'}\,p_1({\bf n\cdot n'}) I({\bf n'},\tau)+
\]
\begin{equation}
\frac{1-\varepsilon_2}{4\pi(1+\eta_{eff})}\int d{\bf n'}\,p_2({\bf n\cdot n'}) I({\bf n'},\tau)  +S_0({\bf n},\tau),
\label{41}
\end{equation} 
\noindent  where the value $\eta_{eff}$ has the form:
\begin{equation}
\eta_{eff}=\frac{\alpha_{1eff}}{\alpha_{2eff}}=\frac{N^{(0)}_1\sigma^{(t)}_1-(\gamma_1+\gamma_3)}{N^{(0)}_2\sigma^{(t)}_2-(\gamma_2+\gamma_3)}.
\label{42}
\end{equation}

Thus, in  stochastic atmosphere with two types of scattering particles the parameter $\eta$ in standard  Eq.(34) takes the value $\eta_{eff}$.  Note that if the particles of the first and the second types are statistically independent one from another, we have $\gamma_3=0$.

 It is very interesting, $\eta_{eff}$ is  longer or less than $\eta$? We consider this problem for specific case, which looks fairly apparent. We take $\gamma_3=0$  and assume, that the correlation lengths $R_1=R_2=R$ and $f_{1\alpha}=f_{2\alpha}=f_{\alpha}$. Besides, we assume that the levels of fluctuations are equal: $\langle N'^2_1\rangle/(N^{(0)}_1)^2 =\langle N'^2_2\rangle/(N^{(0)}_2)^2 \equiv g$. As a result, we obtain:
\begin{equation}
\eta_{eff}\simeq \eta\,\frac{1-gf_{\alpha}\tau_1}{1-gf_{\alpha}\tau_2},
\label{43}
\end{equation}
\noindent where $\tau_1=N^{(0)}_1\sigma^{(t)}_1R$ and  $\tau_2=N^{(0)}_2\sigma^{(t)}_2R$ are the optical depths of correlation length $R$ for the first and second extinction factors, respectively. Thus, $\eta_{eff}>\eta$, if $\tau_2>\tau_1$. In opposite case $\tau_2<\tau_1$ we have $\eta_{eff}<\eta$ . Note that the angular distribution $J({\bf n})$ of emerging radiation depends on parameter $\eta$ (see Ch.5). The difference $\eta_{eff}$ from $\eta$  changes the values  $J({\bf n})$ for stochastic atmosphere from the value in non-stochastic case.

  In the next section we consider the stochastic atmosphere consisting of  free electrons and small anisotropic particles (molecules or dust grains). For simplicity, we will name them as grains. Most interesting examples are the gas dusty accretion discs. Note that according to standard theory with $\alpha_{SS}=0.01$ (Shakura \& Sunyaev 1973; Pariev \& Colgate  2007) the temperature in accretion disc is too high for presence of small dust grains. Recall, that $\alpha_{SS}$ is viscosity factor. It is known that  in AGNs of Seyfert galaxies the optically thick gas dusty tori exist (see, for example, Snedden \& Gaskell 2007;  Gaskell 2011). There exists the dust inflow in accretion disc. This supports the dust component in accretion disc. Of course, instead of free electrons we can consider atoms with the spherical polarizability. In this case we are take $\sigma_T \to \sigma_{atom}$.

\section{ Radiative transfer equation  for Stokes parameters}

The propagation of polarized radiation in stochastic atmosphere is analogous to  the case of intensity propagation. The difference is that instead of the scalar standard radiative transfer equation (Eq.(32)), we consider the matrix transfer equation for Stokes parameters $I({\bf n},r), Q({\bf n},r)$ and $U({\bf n},r)$. Recall, that the Stokes parameter $V({\bf n},r)$ , describing the circular polarization, obeys the separate scalar transfer equation. We do not consider this equation. Chandrasekhar (1960) considered in detail
the system of transfer equations for parameters  $I_l({\bf n},r), I_r({\bf n},r)$ and $U({\bf n},r)$, where intensity $I_l({\bf n},r)$ describes the radiation  linearly polarized in the plane (${\bf nN}$),  and $I_r$ is the  intensity with polarization perpendicular to this plane.  Here ${\bf n}$ is the line of sight and ${\bf N}$ is the normal to the surface of an atmosphere. We introduce parameter $\mu=\cos\vartheta $ , where $\vartheta$ is the angle between ${\bf n}$ and ${\bf N}$. 
 The intensity  $I({\bf n},r)= 
I_l({\bf n},r)+ I_r({\bf n},r)$ and the parameter $Q({\bf n},r)= I_l({\bf n},r)- I_r({\bf n},r)$. Below we consider  the axially symmetric problem, where parameter  $U({\bf n},r)\equiv 0$.  Particularly such problem is the Milne problem in non-conservative medium (optically thick torus or accretion disc in AGN).

 Frequently one  uses the system of equations for   $I({\bf n},r)$ and  $Q({\bf n},r)$. We restrict ourselves by this case. First let us recall the radiative transfer equation for the (column) vector with the components ($I, Q$) in an atmosphere
consisting of averaged small anisotropic particles (molecules or dust grains) and free electrons. The equation  for $I(\mu,r)$  and $Q(\mu,r)$  can  be readily transformed from the equation for the column ($I_l(\mu,r)$,$I_r(\mu,r)$),  presented in Chandrasekhar 1960; Dolginov et al. 1995; Silant'ev et al. 2015 :
\[
 \frac{d}{dr}\left (\begin {array}{c} I(\mu,r) \\ Q(\mu,r) \end{array}\right )
=-\alpha(r) \left (\begin {array}{c} I(\mu,r) \\ Q(\mu,r) \end{array}\right )+\frac{1}{2}\left[\frac{3}{8}\alpha^{(s)}(r)\int^1_{-1}d\mu'\times \right.
\]
\begin{equation}
\left.\left(\begin{array}{rr}3-\mu^2-\mu'^2+3\mu^2\mu'^2,  1-\mu'^2-3\mu^2+3\mu^2\mu'^2\\
 1-\mu^2-3\mu'^2+3\mu^2\mu'^2, 3(1-\mu^2)(1-\mu'^2)\end{array}\right ) +\beta^{(s)}(r)\int_{-1}^1d\mu' \left (\begin{array}{rr}1,  0\\ 0, 0\end{array}\right )\right ]
 \left (\begin {array}{c} I(\mu',r) \\ Q(\mu',r) \end{array}\right )+S(r)\left (\begin {array}{c} 1 \\ 0 \end{array}\right) ,
\label{44}
\end{equation}
\noindent where  the extinction factor is equal to:
\begin{equation}
 \alpha(r)\equiv \alpha_0=N_e^{(0)}(r)\sigma_T+N^{(0)}_g(r)\sigma^{(t)}_g.
\label{45}
\end{equation}
\noindent In Eq.(44) the factors before the integral terms  are:
\begin{equation}
\alpha^{(s)}(r)\equiv \alpha^{(s)}_0=N^{(0)}_e(r)\sigma_T+(1-\varepsilon)N^{(0)}_g(r)\sigma^{(t)}_g\overline b_1
\label{46}
\end{equation}
\begin{equation}
\beta^{(s)}(r)\equiv \beta_0^{(s)}(r)=3(1-\varepsilon)N^{(0)}_g(r)\sigma^{(t)}_g\overline b_2.
\label{47}
\end{equation}

  The values $\sigma^{(s)}_g$ and $\sigma^{(a)}_g$ are the
cross-sections of scattering and absorption by dust grains, $\sigma^{(t)}_g=\sigma^{(s)}_g +\sigma^{(a)}_g$ is the cross-section of total extinction; $\sigma_T$ is the Thomson cross-section. $N_e(r)$ and $N_g(r)$ are the number densities of  free electrons and dust grains, respectively.  The degree of the light absorption
 $\varepsilon=\sigma^{(a)}_g/\sigma^{(t)}_g$, $\mu ={\bf nN}$ is cosine of the angle between the directions of light propagation ${\bf n}$ and the outer normal ${\bf N}$ to plane-parallel semi-infinite atmosphere. 

The first integral term describes the Rayleigh scattering and the second one describes the isotropic scattering by anisotropic part of the dust grains.  Here we have two stochastic values - $N_e(r)=N_e^{(0)}(r) + N'_e(r)$ and $N_g(r)=N_g^{(0)}(r) +N'_g(r)$.

The general consideration and notations,  presented in above section, are also used in this case. We take the  parameter $\eta=N^{(0)}_e\sigma_T/(N^{(0)}_g\sigma^{(t)}_g)$ and $(1-\varepsilon)=\sigma^{(s)}_g/\sigma^{(t)}_g$, i.e. $N_e=N_1$ and $N_g=N_2$, $\varepsilon_1=0$ and $\varepsilon_2\equiv \varepsilon$.
 The  difference  is that  the first integral term depends on  the both types of scatterers. For this reason we  describe the generalized method given in previous chapter once more.

The dimensionless parameters $\overline b_1$ and $\overline b_2$ are related with the anisotropic dust grains (or the anisotropic molecules). Parameter $\overline b_1$ describes the Rayleigh scattering on chaotically oriented dust grains. According to Chandrasekhar (1960) such grains also demonstrate the isotropic nonpolarized scattered  radiation, which is describes by the parameter   
 $\overline b_2$ (see Silant'ev et al. 2017). These parameters obey the relation $\overline b_1 + 3\overline b_2=1$.
 For needle like grains parameters $\overline{b}_1=0.4$ and  $\overline b_2=0.2$.
 For plate like particles  we have $\overline{b}_1=0.7,\, \overline{b}_2=0.1$.
Parameter $\overline{b}_2$ describes the depolarization of radiation, scattered by freely oriented anisotropic
particles. So, the needle like particles depolarize radiation greater than the plate like ones. The relation of parameters  $\overline b_1$ and $\overline b_2$  with the polarizability tensor of a grain (molecule) is given in Dolginov et al. 1995; Silant'ev et al. 2017.

If we take $d\tau=\alpha_0dr$, then Eq.(44) can be written in the form:
\[
 \frac{d}{d\tau}\left (\begin {array}{c} I(\mu,\tau) \\ Q(\mu,\tau) \end{array}\right )=({\bf n}\nabla_{\tau})\left (\begin {array}{c} I(\mu,\tau) \\ Q(\mu,\tau) \end{array}\right )
=- \left (\begin {array}{c} I(\mu,\tau) \\ Q(\mu,\tau) \end{array}\right )+\frac{1}{2}\left [\frac{3}{8}a\int^1_{-1}d\mu'\times \right.
\]
\begin{equation}
\left.\left (\begin{array}{rr}3-\mu^2-\mu'^2+3\mu^2\mu'^2,  1-\mu'^2-3\mu^2+3\mu^2\mu'^2\\
 1-\mu^2-3\mu'^2+3\mu^2\mu'^2, 3(1-\mu^2)(1-\mu'^2)\end{array}\right ) +b\int_{-1}^1d\mu' \left (\begin{array}{rr}1,  0\\ 0, 0\end{array}\right )\right ]
 \left (\begin {array}{c} I(\mu',\tau) \\ Q(\mu',\tau) \end{array}\right )+S(\tau)\left (\begin {array}{c} 1 \\ 0 \end{array}\right) ,
\label{48}
\end{equation}
\noindent where parameters $a$ , $b$  and $\eta$ are:
\begin{equation}
a=\frac{\eta+(1-\varepsilon)\overline b_1}{1+\eta},\quad b=\frac{(1-\varepsilon)3\overline b_2}{1+\eta},
\label{49}
\end{equation}
\begin{equation}
\overline b_1+3\overline b_2=1, \quad a+b=\frac{\eta+1-\varepsilon}{1+\eta},
\label{50}
\end{equation}
\begin{equation}
\eta=\frac{N^{(0)}_e\sigma_T}{N^{(0)}_g\sigma^{(t)}_g}.
\label{51}
\end{equation}

Assuming that all factors in Eq.(44) are the stochastic values, we   average  this equation. The value $\alpha_{eff}$ has the  form:
\begin{equation}
\alpha_{eff}(r)=\alpha_{1eff}+\alpha_{2eff}= [ N^{(0)}_e\sigma_T-(\gamma_+\gamma_3)]+[N^{(0)}_g\sigma^{(t)}_g -(\gamma_2+\gamma_3)], 
\label{52}
\end{equation}
\noindent where
\[\gamma_1=\langle N'^2_e\rangle\sigma_T^2 R_1f_{1\alpha},\quad \gamma_2=\langle N'^2_g\rangle ( \sigma_g^{(t)})^2 R_2 f_{2\alpha},\quad\gamma_3=\langle N'_eN'_g\rangle \sigma_T \sigma^{(t)}_g R_3f_{3\alpha},
\]
\begin{equation}
\alpha_0=N_e^{(0)}\sigma_T+N^{(0)}_g\sigma_g^{(t)},\quad \alpha'=N'_e\sigma_T+N'_g\sigma_g^{(t)},
\label{53}
\end{equation}
\noindent In Eq.(52) three types of the averages exist - $\langle N'_e(s)N'_e(s'')\rangle$, $\langle N'_g(s')N'_g(s'')\rangle$ and $\langle N'_e(s')N'_g(s'')\rangle$. Every of them is analogous to Eq.(25). As in previous section, we introduced different correlation functions and different lengths of correlation. 

Averaging the value $\alpha^{(s)}$, we obtain:
\begin{equation}
\alpha^{(s)}_{eff}=\alpha_{1eff}+(1-\varepsilon)\alpha_{2eff}\overline b_1 .
\label{54}
\end{equation}
\begin{equation}
\alpha^{(s)}_0= N^{(0)}_e\sigma_T+(1-\varepsilon)\overline b_1 N^{(0)}_g\sigma^{(t)}_g,\quad \alpha'_s=N'_e\sigma_T+(1-\varepsilon)\overline b_1N'_g\sigma^{(t)}_g.
\label{55}
\end{equation}

For the second integral term in Eq.(44) we have:
\begin{equation}
\beta^{(s)}_{eff}=3\overline b_2(1-\varepsilon)[N^{(0)}_g\sigma^{(t)}_g-(\gamma_2+\gamma_3)]\equiv3\overline b_2(1-\varepsilon)\alpha_{2eff},
\label{56}
\end{equation}

\begin{equation}
\beta^{(s)}_0=3\overline b_2 N_g^{(0)}\sigma_g^{(s)}=3\overline b_2 (1-\varepsilon)N^{(0)}_g\sigma^{(t)}_g,\quad \beta'_s=3\overline b_2 (1-\varepsilon) N'_g\sigma^{(t)}_g.
\label{57}
\end{equation}

If we introduce the effective optical depth $d\tau=\alpha_{eff}dr$, then  the equation for averaged values $I_0(\mu,r)$ and  
$Q_0(\mu,r)$  takes the form analogous to Eq.(48), where $a\to a_{eff}$ and $b\to b_{eff}$:
\[
 \frac{d}{d\tau}\left (\begin {array}{c} I_0(\mu,\tau) \\ Q_0(\mu,\tau) \end{array}\right )=({\bf n}\nabla_{\tau})\left (\begin {array}{c} I_0(\mu,\tau) \\ Q_0(\mu,\tau) \end{array}\right )
=- \left (\begin {array}{c} I_0(\mu,\tau) \\ Q_0(\mu,\tau) \end{array}\right )+\frac{1}{2}\left [\frac{3}{8}a_{eff}\int^1_{-1}d\mu'\times \right.
\]
\begin{equation}
\left.\left (\begin{array}{rr}3-\mu^2-\mu'^2+3\mu^2\mu'^2,  1-\mu'^2-3\mu^2+3\mu^2\mu'^2\\
 1-\mu^2-3\mu'^2+3\mu^2\mu'^2, 3(1-\mu^2)(1-\mu'^2)\end{array}\right ) +b_{eff}\int_{-1}^1d\mu' \left (\begin{array}{rr}1,  0\\ 0, 0\end{array}\right )\right ]
 \left (\begin {array}{c} I_0(\mu',\tau) \\ Q_0(\mu',\tau) \end{array}\right )+S_0(\tau)\left (\begin {array}{c} 1 \\ 0 \end{array}\right ) ,
\label{58}
\end{equation}
\noindent where we  introduced the notations:

\begin{equation}
a_{eff}=\frac{\alpha^{(s)}_{eff}}{\alpha_{eff}}=\frac{\eta_{eff}+(1-\varepsilon)\overline b_1}{1+\eta_{eff}},\quad b_{eff}=\frac{\beta^{(s)}_{eff}}{\alpha_{eff}}=\frac{(1-\varepsilon)3\overline b_2}{1+\eta_{eff}}, \quad a_{eff}+b_{eff}=\frac{\eta+1-\varepsilon}{1+\eta}.
\label{59}
\end{equation}
\noindent Recall, that $\overline b_1+3\overline b_2=1$.

 Parameter $\eta_{eff}$ is equal to:
\begin{equation}
\eta_{eff}=\frac{N^{(0)}_e\sigma_T-(\gamma_1+\gamma_3)}{N^{(0)}_g\sigma^{(t)}_g - (\gamma_2+\gamma_3)}.
\label{60}
\end{equation}
\noindent  This formula coincides with Eq.(42). The estimates in Eq.(43) take place also for Eq.(60). Recall, that here $\tau_1=N_e\sigma_T R$ and   $\tau_2=N_g\sigma_g^{(t)} R$.
 Eq.(58) can be written in more simple form in the new notations:
\begin{equation}
W_{eff}=\frac{a_{eff}}{a_{eff}+b_{eff}},\quad C_{eff}=\frac{W_{eff}}{8}.
\label{61}
\end{equation}
\noindent  Using the new factorization (see Frisch 2017), Eq.(58) can be presented as:
\begin{equation}
 \frac{d}{d\tau}\left (\begin {array}{c} I_0(\mu,\tau) \\ Q_0(\mu,\tau) \end{array}\right )=({\bf n}\nabla_{\tau})\left (\begin {array}{c} I_0(\mu,\tau) \\ Q_0(\mu,\tau) \end{array}\right) 
=- \left (\begin {array}{c} I_0(\mu,\tau) \\ Q_0(\mu,\tau) \end{array}\right )+\hat A(\mu){\bf K}(\tau)+ S_0(\tau)\left (\begin {array}{c} 1 \\ 0 \end{array}\right) ,
\label{62}
\end{equation}
\noindent where we introduced the vector ${\bf K}(\tau)$:

\begin{equation}
{\bf K}(\tau)=\frac{1}{2}\int^1_{-1}d\mu' \hat A^T(\mu') \left(\begin {array}{c} I_0(\mu',\tau) \\ Q_0(\mu',\tau) \end{array}\right ).
\label{63}
\end{equation}
\noindent  The factorization matrix $\hat A(\mu)$ has the form: 
\begin{equation}
\hat A(\mu)= \sqrt{(a_{eff}+b_{eff})}\left (\begin{array}{rr}1 ,\,\, \,\sqrt{C_{eff}}(1-3\mu^2) \\ 0 ,\,\,\,3\sqrt{C_{eff}} (1-\mu^2) \end{array}\right).
\label{64}
\end{equation}
\noindent Superscript $T$ stands for the matrix transpose.    It is easy prove that the conservation law for the radiation flux (19) is also valid from Eqs.(44),(48),(58) and (62) for $\varepsilon =0$ and $S(\tau)=0$, if we take into account the relation $\overline b_1+3\overline b_2=1$.

The vector ${\bf K}(\tau)$ obeys the integral equation (see Silant'ev et al. 2015, 2017):
\begin{equation}
{\bf K}(\tau)={\bf g}(\tau)+\int_0^{\infty}d\tau' \,\hat L(|\tau-\tau'|){\bf K}(\tau').
\label{65}
\end{equation}
\noindent  The term ${\bf g}(\tau)$ presents the contribution of the source $S_0(\tau)$. The matrix kernel $\hat L(|\tau-\tau'|)$ has the form:
\begin{equation}
\hat L(|\tau-\tau'|)=\int_0^1\frac{d\mu}{\mu}\exp{\left(-\frac{|\tau-\tau'|}{\mu}\right)}\hat \Psi(\mu),
\label{66}
\end{equation}
\noindent where
\begin{equation}
\hat \Psi(\mu)=\frac{1}{2}\hat A^T(\mu)\hat A(\mu).
\label{67}
\end{equation}
\noindent It should be noted that Eq.(65) can be solved by resolvent matrix technique (see Silant'ev et al. 2014, 2015).

In the limiting cases of  Eq.(62), where  $N_g=0$ (the scattering on free electrons) and  $N_e=0$ (the scattering on small dust grains) this  equation  formally coincides with that in non-stochastic atmosphere. 
 The difference is that in stochastic atmosphere we have $d\tau=\alpha_{eff}dr$ instead of $d\tau=\alpha_0dr$ for
non-stochastic case.

\section {The generalized Milne problem in stochastic atmosphere}

The standard Milne problem describes the propagation of non-polarized radiation from deep layers of non-absorbed atmosphere.
In such case the flux ${\bf F}(\tau)$ is similar at every distance in the atmosphere. The generalized Milne problem describes the
propagation of non-polarized radiation from deep layers of absorbing atmosphere. Clearly, the radiation flux ${\bf F}(\tau)$ in this
case is different at different distances in the atmosphere. The radiation flux emerging from atmosphere can be deal with the observing flux. In the both cases of the Milne problems we can calculate the angular distribution  $J(\mu) $ and the degree of polarization $p(\mu)$ for emerging radiation.   The exact solution of generalized Milne problem for Eq.(48) was given in Silant'ev et al. 2017.  In that paper we presented the angular distribution $J(\mu)$ and polarization degree $p(\mu)$ for various values of absorption degree  $\varepsilon$ and parameter $\eta$. The case of pure electron atmosphere was also presented.

The specific feature of the Milne problems is that we are to solve the integral equation for ${\bf K}(\tau)$ without the source term ${\bf S}(\tau)$, i.e. this is the homogeneous integral equation (see Sobolev 1969, Silant'ev et al. 2017).  It is known (see Smirnov 1964) that such equations have the nonzero solution, if there is the solution of characteristic equation  for number $k$($0 \le k\le1$):
\begin{equation}
\left|\left(\hat{E}-2\int_0^1 d\mu\frac{\hat{\Psi}(\mu)}{1- k^2\mu^2}\right)\right|=0.
\label{68}
\end{equation}
\noindent  Note that the matrix $\hat \Psi(\mu)$ depends on $a_{eff}$ and $b_{eff}$.
 The most important feature of the Milne problems is that
the angular distribution of emerging intensity $J(\mu)$ depends  on $k$ as:
\begin{equation}
J(\mu)\sim \frac{1}{1-k\mu},
\label{69}
\end{equation}
\noindent i.e. the angular distribution peaks along the normal ${\bf N}$ to the  atmosphere.
  
   Equation (58), describing the radiative transfer in stochastic atmosphere, coincides with Eq.(48), if we  take $a\to  a_{eff}$ and $b\to b_{eff}$.  Thus, the angular distribution $J(\mu)$ and polarization degree $p(\mu)$,  given in Silant'ev et al.(2017), are true for Eq.(58), if  instead of parameters $a$ and $b$ we take the parameters $a_{eff}$ and $b_{eff}$. If the observed values $J(\mu)$ and $p(\mu)$ do not correspond to accepted parameters $\alpha$ and $\eta$, then we can take the values $\alpha_{eff}$ and $\eta_{eff}$ in order to obtain the observed data. It should be noted that the parameter $\eta_{eff}$ may be both longer and less of parameter $\eta$.
 
It is known (see Silant'ev et al. 2017) that for Eq.(48) the approximate formula for characteristic number $k$ has the form:

\begin{equation}
k\simeq \sqrt{\frac{3\varepsilon}{1+\eta}},\quad \varepsilon\ll 1.
\label{70}
\end{equation}
\noindent   For stochastic atmosphere (Eq.(58)) the expression (70) transforms to:
\begin{equation}
k\simeq \sqrt{\frac{3\varepsilon}{1+\eta_{eff}}}.
\label{73}
\end{equation}
\noindent  Thus, in stochastic atmosphere the angular distribution $J(\mu)$ and polarization degree $p(\mu)$ differ from that in non-stochastic one.
It is interesting to note that formulas (70) and (71) are independent of the form of dust grains. Silant'ev et al.(2017)  demonstrated that the characteristic number $k$ depends very little on the form of dust grains only for $\varepsilon \simeq 0.5$.

\section{Conclusion}

  The objective of the paper is to show how  the radiative transfer equations are changed  in stochastic atmospheres.
In this paper we considered the influence of Gaussian  fluctuations of the number densities of scattering particles on the radiative transfer process. The atmosphere is taken non-magnetized.  We  considered the radiative transfer equation in the atmosphere consisting of both one type scattering particles  and two types ones. In both cases the difference of new equation  from the standard  one  is that the optical depth $d\tau=\alpha dr$ is substituted by new dimensionless optical depth $d\tau=\alpha_{eff}dr$. It is found that $\alpha_{eff}<\alpha$, i.e. the stochastic atmosphere is more transparent than the non-stochastic one. For the atmosphere consisting of two types of scattering  particles the difference is more significant. In addition to $d\tau \to \alpha_{eff}dr$, the new factors arise before the integral terms. This factors  depend on the correlation functions of number densities $\langle N'_1(r')N'_1(r'')\rangle$,  $\langle N'_2(r')N'_2(r'')\rangle$ and  $\langle N'_1(r')N'_2(r'')\rangle$, where $N'_1$ and $N'_2$ are the number densities of the first and second types scattering particles, respectively. Note that the integral terms of standard equation for two types scattering particles  depend  on one dimensionless parameter $\eta=N_1\sigma^{(t)}_1/(N_2\sigma^{(t)}_2)$, where $\sigma^{(t)}_1 $ and  $\sigma^{(t)}_2 $ are the total cross-sections of the first and the second types of scattering particles, respectively. In stochastic atmosphere the $\eta$ - parameter transforms to $\eta_{eff}$ - parameter,  which can be both longer and less than the parameter $\eta$.  This  means that the results, obtained for standard radiative transfer equation, can be used, if we substitute the coefficients $\alpha$ and $\eta$ by the factors $\alpha_{eff}$ and $\eta_{eff}$. We give the estimates of parameter $\eta_{eff}$ as a function of the level of fluctuations and optical depths of correlation lengths.

\section{\bf Acknowledgements.}
This research was supported by the Basic Research Program N 21 of Presidium of Russian Academy of Sciences, 
the Program  of the Department of Physical Sciences of Russian
Academy of Sciences No 2 and the President  Program "The leading scientific schools" N 7241. 

 The authors are very grateful to a referee for many useful remarks and advices.

\end{document}